\documentclass[]{article}
\usepackage{times}
\usepackage{latexsym}
\usepackage{amsbsy}
\usepackage{ifthen}

\usepackage[dvips]{epsfig}

\begin{document}


\def\##1{\underline #1}
\def\=#1{\underline{\underline #1}}

\def\eps{\epsilon}
\def\epso{\epsilon_0}
\def\muo{\mu_0}
\def\ko{k_0}
\def\kosq{k_0^2}
\def\lambdao{\lambda_0}
\def\etao{\eta_0}
\def\.{\mbox{ \tiny{$^\bullet$} }}

\def\curl{\nabla\times}
\def\div{\nabla \mbox{ \tiny{$^\bullet$} }}

\def\epsa{\epsilon_a}
\def\epsb{\epsilon_b}
\def\alfaa{\alpha_a}
\def\alfab{\alpha_b}

\def\rp{r_\parallel}
\def\rs{r_\perp}

\def\ux{\#{u}_x}
\def\uy{\#{u}_y}
\def\uz{\#{u}_z}
\def\up{\#{u}_+}
\def\um{\#{u}_-}

\def\le{\left(}
\def\ri{\right)}
\def\les{\left[}
\def\ris{\right]}
\def\lec{\left\{}
\def\ric{\right\}}

\def\c#1{\cite{#1}}
\def\l#1{\label{#1}}
\def\r#1{(\ref{#1})}

\newcommand{\mat}[1] {\left[\begin{array}{cccc}
            #1 \end{array}\right]}

\newcommand{\mb}[1]{\mbox{\boldmath$\bf#1$}}


\noindent  \emph{\Large Positive and Negative Goos--H\"anchen Shifts and Negative Phase--Velocity Mediums (alias Left--Handed
Materials)}\\

\noindent{\b Akhlesh Lakhtakia},
 CATMAS---Computational \& Theoretical
Materials Science Group, Department of Engineering Science and
Mechanics, Pennsylvania State University, University Park, PA
16802--6812, USA.\\ E--mail: AXL4@psu.edu\\

\emph{On total reflection from a half--space filled
with an isotropic, homogeneous, weakly dissipative,
 dielectric--magnetic medium with negative
phase--velocity (NPV) characteristics, it is shown here
that a linearly polarized beam can
experience either a negative or a positive Goos--H\"anchen shift.
The sign of the shift depends on the polarization state of the beam
as well as on the signs of the real parts of the  permittivity and
the permeability of the NPV medium.}


\section{Introduction}
This communication addresses the topic of Goos--H\"anchen shifts of beams
at the specularly flat interface of two isotropic dielectric--magnetic mediums, one
of which
displays negative phase--velocity (NPV) characteristics and the other displays positive
phase--velocity (PPV) characteristics. While positive phase velocities are
commonly encountered
\c{Chen}, negative phase velocities are permitted by the
structure of the Maxwell postulates and several instances have been
recorded \c{Pen03}.

{\em Negative Phase Velocity:\/} Isotropic dielectric--magnetic 
materials with negative phase
velocity~---~i.e., phase velocity opposed in direction to the time--averaged
Poynting vector~---~have attracted much attention of late \c{LMWaeu}.
Over three decades ago, Veselago \cite{Ves} suggested many unusual
properties of  materials with negative real permittivity
{\em and\/} negative real permeability at a certain frequency,
including inverse refraction, negative radiation pressure,
and 
inverse
Doppler effect. But his suggestion was completely speculative
until a breakthrough was announced by Smith {\em et al.\/}
\cite{Schultz1} in 2000. Many names have been
proposed for these materials \c{LMWaeu}~---~of which
the most inappropriate is  {\em left--handed
material\/}. The least ambiguous of all extant names is  {\em NPV
material\/}.
Two recent publications are recommended for ongoing developments
on NPV materials \c{LMWaeu,OptExp}.

\begin{figure}[!ht]
\centering \psfull
\epsfig{file=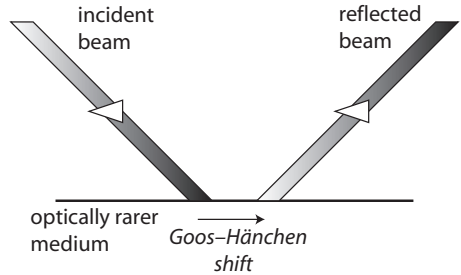,width=6cm }
\caption{Schematic of the {\em positive} Goos--H\"anchen shift on total reflection.
}
\label{posGH}
\end{figure}

\begin{figure}[!ht]
\centering \psfull
\epsfig{file=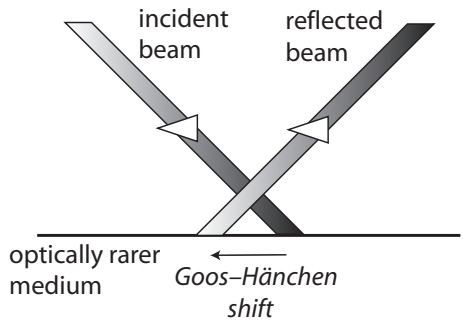,width=6cm }
\caption{Schematic of the {\em negative} Goos--H\"anchen shift on total reflection.
}
\label{negGH}
\end{figure}

{\em Goos--H\"anchen Shift:\/} If a beam of light were to impinge on a
planar interface with an optically rarer dielectric
material and total reflection were
to occur, Newton had conjectured that 
 the reflected beam would be displaced forward by a distance $d$
parallel to the interface \c{Lotsch1}; see Figure \ref{posGH}.
Performing an ingenious experiment some two centuries later, 
Goos and H\"anchen were able to prove Newton correct \c{GH1,GH2}.
Since then, 
these shifts have been estimated as well as measured for planar
interfaces between several different pairs
of homogeneous materials \c{HNS,DB96}. Although Figure \ref{posGH} shows a positive 
shift, negative shifts are also possible \c{AFM78}~---~without violation of causality~---~
as shown in Figure \ref{negGH}.
The significance of Goos--H\"anchen shifts has grown with the
emergence of   near--field optical microscopy
and lithography \c{Forn}.

The vast majority of publications on Goos--H\"anchen shifts deal
with dielectric mediums. The role of permeability appears to have
been largely ignored. But permeability is virtually of the same
status as permittivity for NPV materials.

Reversal
of the signs of the real parts of both the permittivity and the
permeability of the optically rarer medium has been proved
to result  in the reversal of the sign (and, therefore, the direction)
of the Goos--H\"anchen shift \c{Lem}. Furthermore, if the optically rarer medium
has negative real permittivity {\em and\/} negative real permeability,
the Goos--H\"anchen shifts of both perpendicularly and parallel
polarized beams have been shown to be negative \c{Ber}. However,
a NPV material need not have both of those constitutive
quantities as negative \c{MLW02},
which prompted the research reported here.

\section{Theoretical preliminaries}
Consider the planar interface of two
homogeneous mediums labeled $a$ and $b$,
with respective relative permittivities $\eps_{a,b}=\eps_{a,b}'+i\eps_{a,b}''$ and 
relative permeabilities $\mu_{a,b}=\mu_{a,b}'+i\mu_{a,b}''$ at the angular
frequency $\omega$ of interest.
Medium $a$ is supposed to be
nondissipative (i.e., $\eps_{a}''=0$ and $\mu_{a}''=0$),
while dissipation in medium $b$ is assumed to be very small
(i.e., $\eps_b''\ll\eps_b'$ and $\mu_b''\ll\mu_b'$) for the sake of simplicity.
The incident and the reflected beams lie in medium $a$, and
an $\exp(-i\omega t)$ time--dependence is implicit.

The planewave reflection coefficients for this situation are given
by 
\begin{equation}
\label{rp}
\rp = \frac{1-(\alfaa/\alfab)(\epsb/\epsa)}{1+(\alfaa/\alfab)(\epsb/\epsa)}
\end{equation}
for parallel polarization, and
\begin{equation}
\label{rs}
\rs = -\,\frac{1-(\alfaa/\alfab)(\mu_b/\mu_a)}{1+(\alfaa/\alfab)(\mu_b/\mu_a)}\,
\end{equation}
for perpendicular polarization. In these equations,
\begin{equation}
\alpha_{a,b} = \ko\sqrt{\epsilon_{a,b}\mu_{a,b}-\epsa\mu_a\sin^2\theta_{inc}}
\end{equation}
involve the free--space wavenumber $\ko$ and the angle of incidence
$\theta_{inc} \in \les0,\pi/2\ri$. We must choose ${\rm Im}\les\alpha_{b}\ris \geq 0$,
as befits any passive medium.

The phase velocity vector opposes the
direction of the time--averaged Poynting vector in medium
$b$, whenever the inequality
\begin{equation}
\left( \big\vert\eps_b\big\vert - {\eps'_b} \right) \left( \big\vert\mu_b\big\vert- {\mu'_b} \right)
> \eps''_b\mu''_b
\label{inequality}
\end{equation}
holds \cite{MLW02}. 
Thus, the simultaneous satisfaction of
 both $\eps'_b <0$ and
$\mu'_b<0$ is a sufficient, but not necessary, requirement  for
the phase velocity to be negative. In other
words, both $\eps'_b$ and
$\mu'_b$ do not have to be negative for the phase
velocity to be negative.

\section{Analysis and conclusions}

When  investigating Goos--H\"anchen shifts, it is best to write the
reflection coefficients as
\begin{equation}
r_{\parallel,\perp} = \big\vert r_{\parallel,\perp}\big\vert\,\exp\le i\varphi_{\parallel,\perp}\ri
\,.
\end{equation}
As dissipation in medium $b$ is assumed here to be
very small, the reflection coefficients
are essentially of unit amplitude when $\theta_{inc}$ exceeds the critical angle.
Artmann \c{Artmann} showed that Goos--H\"anchen shifts can then be
be estimated as
\begin{equation}
\label{ArtmannE}
d_\parallel=-\,\frac{\partial \varphi_\parallel}{\partial \kappa}\,,\qquad
d_\perp=-\,\frac{\partial \varphi_\perp}{\partial \kappa}\,,
\end{equation}
where $\kappa =\ko \sqrt{\eps_a'\mu_a'}\,\sin\theta_{inc}$. The foregoing
expressions are not adequate when $\theta_{inc}$ is very close
to either the critical angle or $\pi/2$, but suffice for the present purpose.

With $\eps_{a,b}$,  $\mu_{a,b}$  and  $\alpha_a$ real--valued,
$\alfab$ is purely imaginary when the angle of incidence
exceeds the critical angle. Then, the expressions
\begin{equation}
\varphi_\parallel=-2\tan^{-1}\le \frac{\alfaa}{\vert\alfab\vert}\,\frac{\epsb}{\epsa}\ri
\end{equation}
and
\begin{equation}
\varphi_\perp=\pi-2\tan^{-1}\le \frac{\alfaa}{\vert\alfab\vert}\,\frac{\mu_b}{\mu_a}\ri
\end{equation} 
follow from \r{rp} and \r{rs}.
Reversal of the sign of $\epsb$ would certainly affect $\vert\alfab\vert$,
but a more significant effect shall be on the sign of $\varphi_\parallel$ (in contrast
to $\varphi_\perp$) and therefore on the sign of $d_\parallel$. Likewise,
reversal of the sign of $\mu_b$ would change the signs of $\varphi_\perp$ and  $d_\perp$.
This understanding should hold also when medium $b$ is weakly
dissipative (i.e., $\eps_b''\ll\eps_b'$ and $\mu_b''\ll\mu_b'$).

We can conclude the following: 
Let the medium of incidence and reflection be nondissipative
and of the PPV type,
and the refracting medium be optically rarer as well as of the NPV type
with weak dissipation.
When the conditions for
total reflection prevail, the Artmann expressions \r{ArtmannE} suggest that negative Goos--H\"anchen
shifts are possible for 
\begin{itemize}
\item only parallel--polarized beams if  
 the real part of the  permittivity of the NPV medium is
negative,
\item only perpendicularly polarized beams if the real part of the
permeability of the NPV medium is
negative, and
\item both parallel-- and perpendicularly polarized beams if the real parts of both
the permittivity and the permeability of the NPV medium are negative.
\end{itemize}






\end{document}